\documentclass[prb,aps,amssymb,showpacs,twocolumn]{revtex4}
\usepackage{amsmath}
\usepackage{amssymb}
\usepackage{amsthm}
\usepackage{amsfonts}
\usepackage{enumerate}
\usepackage{latexsym}
\usepackage{bm}
\usepackage{graphicx}
\usepackage{subfigure}

\newcommand{\beq}{\begin{equation}}
\newcommand{\eneq}{\end{equation}}

\input{epsf}

\begin{document}

\tolerance 10000

\newcommand{\vk}{{\bf k}}

%\draft

\title{Generalized Clustering Conditions of Jack Polynomials at Negative Jack Parameter $\alpha$}

\author{B. Andrei Bernevig$^{1,2}$ and F.D.M. Haldane$^2$}

\affiliation{ } \affiliation{$^1$Princeton Center for Theoretical
Physics, Princeton, NJ 08544} \affiliation{$^2$ Department of
Physics, Princeton University, Princeton, NJ 08544}

\begin{abstract}
We present several conjectures on the behavior and clustering
properties of Jack polynomials at \emph{negative} parameter
$\alpha=-\frac{k+1}{r-1}$, of partitions that violate the $(k,r,N)$
admissibility rule of Feigin \emph{et. al.}
[\onlinecite{feigin2002}]. We find that "highest weight" Jack
polynomials of specific partitions represent the minimum degree
polynomials in $N$ variables that vanish when $s$ distinct clusters
of $k+1$ particles are formed, with $s$ and $k$ positive integers.
Explicit counting formulas are conjectured. The generalized
clustering conditions are useful in a forthcoming description of
fractional quantum Hall quasiparticles.
\end{abstract}

\date{\today}


\maketitle

\section{Introduction}

The Jack polynomials are a family of symmetric homogenous
multivariate polynomials characterized by a dominant partition
$\lambda$ and a rational number parameter $\alpha$. They appeared in
physics in the context of the Calogero-Sutherland model
[\onlinecite{sutherland1971}] for positive coupling $\alpha$.

In a recent paper, Feigin \emph{et. al.} [\onlinecite{feigin2002}]
initiated the study of Jack polynomials (Jacks) in $N$ variables, at
negative rational parameter $\alpha_{k,r} = -\frac{k+1}{r-1}$
($k+1$, $r-1$ relatively prime), of certain ($k$,$r$,$N$)-admissible
partitions $\lambda:$ $\lambda_i - \lambda_{i+k} \ge r$, by proving
they form a basis of a differential ideal $I_N^{k,r}$ in the space
of symmetric polynomials. They showed that the set of Jacks with
parameter $\alpha_{k,2}$ and ($k$,$2$,$N$)-admissible $\lambda$ are
a basis for the space of symmetric homogeneous polynomials that
vanish when $k+1$ variables $z_i$ coincide. In
[\onlinecite{bernevig2007}], we found that these Jacks naturally
implement a type of ``generalized Pauli principle'' on a
generalization of Fock spaces for abelian and non-abelian fractional
statistics [\onlinecite{haldaneUCSB2006}]. We found that (bosonic)
Laughlin, Moore-Read, and Read-Rezayi Fractional Quantum Hall (FQH)
wavefunctions (as well as others, such as the state Simon \textit{et
al.} [\onlinecite{simon2006}] have called the ``Gaffnian'') can be
explicitly written as \emph{single} Jack symmetric polynomials. We
identified $r$ as the minimum power (cluster angular momentum) with
which the admissible Jacks vanish as a cluster of $k+1$ particles
come together.

In [\onlinecite{bernevig2007}] we adopted a physical perspective to
the Jack problem and uniquely obtained the abelian and non-abelian
FQH ground states and the admissibility rule on partitions by
imposing, on an arbitrary Jack polynomial, a highest and lowest
weight condition common in FQH studies on the sphere
[\onlinecite{haldane1983}]. However, in imposing only the highest
weight condition on the Jacks we obtained another infinite series of
polynomials of partitions which violate the Feigin \emph{et. al.}
admissibility rule. The $(k,r,N)$-admissible configurations of
[\onlinecite{feigin2002}] do not exhaust the space of well-behaved
Jack polynomials at negative $\alpha_{k,r}$. We find an infinite
series of Jack polynomials, with partitions characterized by a
different integer $s$ which are still well-behaved at negative
rational $\alpha_{k,r}$. This paper is devoted to analyzing the
clustering properties and the counting of such polynomials. We note
that these polynomials can be interpreted as lowest Landau level
(LLL) many-body wavefunctions and have applications in the
construction of FQH quasiparticle excitations
[\onlinecite{uslater}].

We obtain a characterization of the symmetric polynomials in $N$
variables $P(z_1,z_2,...,z_N)$ satisfying the following set of
generalized clustering (vanishing) conditions: (i)
$P(z_1,z_2,...,z_N)$ vanishes when we form $s$ distinct clusters,
each of $k+1$ particles, but remains finite when $s-1$ distinct
clusters of $k+1$ particles are formed and (ii) $P(z_1,z_2,...,z_N)$
does not vanish when a large cluster of $s(k+1)-1$ particles is
formed. $s$ and $k$ are integers greater or equal to $1$ and the
case $s=1$ is the case described by Feigin \emph{et. al.}
[\onlinecite{feigin2002}]. More precisely, let $F$ be the space of
all polynomials satisfying the clustering condition:
\begin{widetext}
\begin{equation}
P(z_1=...=z_{k+1},z_{k+2}=...=z_{2(k+1)},...,z_{(s-1)(k+1)+1}=...=z_{s(k+1)},
z_{s(k+1)+1},z_{s(k+1)+2},...,z_N) =0 \label{clustering01}
\end{equation}
\end{widetext}
\noindent  Let $F_1$ be the space of all polynomials satisfying both
Eq.(\ref{clustering01}), \emph{and} the clustering condition
$P(z_1=...=z_{s(k+1)-1}, z_{s(k+1)},z_{s(k+1)+1},...,z_N) = 0$. In
this paper we look at the coset space $F/F_1$ and focus on two
problems: $(1)$ We find the generators of the ideal above, which are
the \emph{minimum degree} polynomials (for $N$ particles) satisfying
the clustering conditions above, and $(2)$ We give a
(\emph{conjectured}) analytic expression for the number of linearly
independent polynomials in $N$ variables, of momentum (total degree
of the polynomial) $M$ and of flux (maximum power in each variable)
$N_\Phi$ that span the coset $F/F_1$. Our motivation is to find new
properties of Jack polynomials at \emph{negative} $\alpha$ which are
applicable to the study of FQH quasiparticles
[\onlinecite{uslater}]. Our results are also related to the
Cayley-Sylvester problem of coincident loci
[\onlinecite{kasatani2004,chipalkatti2003}].

\section{Properties of Jack Polynomials}

\begin{figure}
\includegraphics[width=3.2in, height=1.9in]{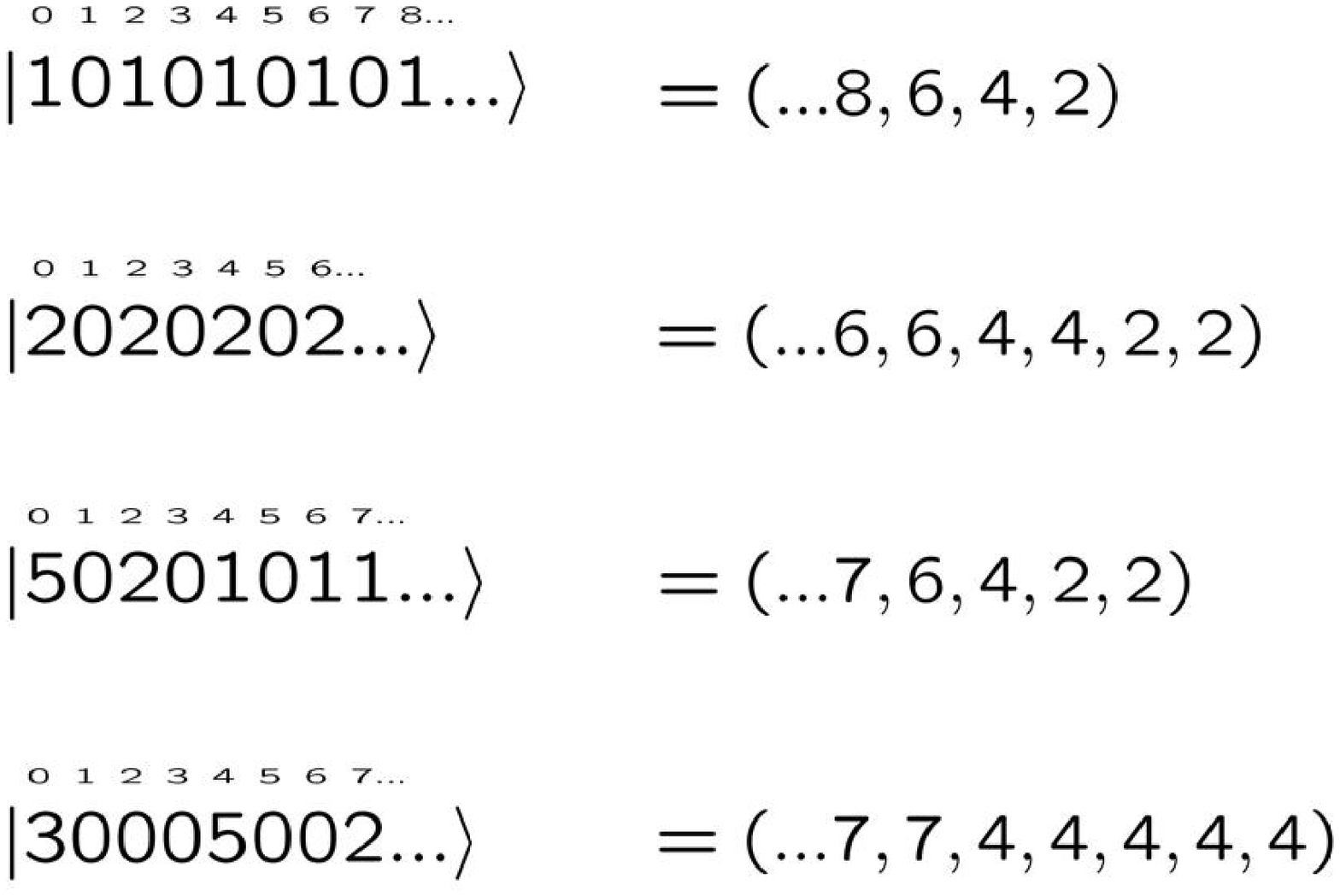}
\caption{Examples of occupation to monomial basis conversion
$n(\lambda) \rightarrow \lambda$}\label{occupation}
\end{figure}

The Jacks $J^{\alpha}_{\lambda}(z)$ are symmetric homogeneous
polynomials in $z$ $\equiv$ $\{z_1,z_2,\ldots ,z_N\}$, labeled by a
partition $\lambda$ with length $\ell_{\lambda} \le N$, and a
parameter $\alpha$; the partition $\lambda$ can be represented as a
(bosonic) occupation-number configuration $n(\lambda)$ =
$\{n_m(\lambda),m=0,1,2,\ldots\}$ of each of the lowest Landau level
(LLL) orbitals with angular momentum $L_z = m \hbar$ (see
Fig[\ref{occupation}]), where, for $m > 0$, $n_m(\lambda)$ is the
multiplicity of $m$ in $\lambda$. When $\alpha$ $\rightarrow$
$\infty$, $J^{\alpha}_{\lambda}$ $\rightarrow$ $m_{\lambda}$, which
is the monomial wavefunction of the free boson state with
occupation-number configuration $n(\lambda)$; a key property of the
Jack $J^{\alpha}_{\lambda}$ is that its expansion in terms of
monomials only contains terms $m_{\mu}$ with $\mu$ $\le$ $\lambda$,
where $\mu$ $<$ $\lambda$ means the partition $\mu$ is
\textit{dominated} by $\lambda$[\onlinecite{stanley1989}]. Jacks are
also eigenstates of a Laplace-Beltrami operator $\mathcal H_{\rm
LB}(\alpha)$ given by
\begin{equation}
\sum_i \left ( z_i \frac{\partial}{\partial z_i} \right )^2 +
\frac{1}{\alpha}\sum_{i<j}\frac{z_i+z_j}{z_i-z_j} \left ( z_i
\frac{\partial}{\partial z_i} - z_j \frac{\partial}{\partial z_j}
\right ).
\end{equation}
\noindent A partition $\lambda$ is
``($k$,$r$,$N$)-admissible''[\onlinecite{feigin2002}] if
$n(\lambda)$ obeys a ``generalized Pauli principle'' where, for all
$m\ge 0$, $\sum_{j=1}^rn_{m+j-1}$ $\le$ $k$, so $r$ consecutive
``orbitals'' contain no more than $k$ particles
[\onlinecite{bernevig2007}].

Partitions $\lambda$ can be classified by $\lambda_1$, their largest
part.  When $J^{\alpha}_{\lambda}$ is expanded in occupation-number
states (monomials), no orbital with $m >\lambda_1$ is occupied, and
Jacks with $\lambda_1 \le N_{\Phi}$ form a basis of FQH states on a
sphere surrounding a monopole with charge
$N_{\Phi}$[\onlinecite{haldane1983}].   Uniform states on the sphere
satisfy the conditions $L^+\psi = 0$ (highest weight, HW) and
$L^-\psi = 0$ (lowest weight, LW) where:
\begin{displaymath}
L^+ = E_0; \;\;\; L^- = N_{\Phi}Z-E_2; \;\;\; L^z =  \frac{1}{2} N
N_\Phi -E_1
\end{displaymath}
\begin{equation}
E_n = \sum_iz_i^n \frac{\partial}{\partial z_i}
\end{equation}
\noindent where $Z$ $\equiv$ $\sum_i z_i$. When both conditions are
satisfied, $E_1\psi$ $\equiv$ $M\psi$ = $ \frac{1}{2}
NN_{\Phi}\psi$. The $L^+,L^-,L^z$ operators endow the polynomial
space with an angular momentum structure which we use to
characterize the polynomials. Any homogeneous polynomial is an
eigenstate of the $L^z$ operator; let the $L^z$ eigenvalue of the HW
Jacks be $l^{max}_z$. The HW states then have $\vec{L}^2=
\frac{1}{2}(L^+ L^- + L^- L^+) + L^zL^z = l^{max}_z(l^{max}_z+1)$,
and hence the HW polynomials are the $(l, l_z)= (l^{max}_z,
l^{max}_z)$ states of a $2 l^{max}_z +1$ angular momentum multiplet
of linearly independent polynomials. The non-HW states of $(l, l_z)
= (l^{max}_z, l^{max}_z - i)$, with $i=1,...,2 l^{max}_z$ can be
obtained by successive application of the lowering operator
$(L^-)^i$ on the HW states. They are linearly independent by virtue
of having the same $N_\Phi$ but different total degree $M$. Applying
the $L^-$ operator $2 l^{max}_z+1$ times kills the state. This
angular momentum structure is extensively used in studies of FQH
states on the sphere and we find it to be extremely valuable in the
empirical polynomial counting presented in the following sections.

It is very instructive to find the conditions for a Jack to satisfy
the HW condition, $E_0J^{\alpha}_{\lambda}$ = 0. The action of $
E_0$ on a Jack can be obtained from a formula due to Lassalle
[\onlinecite{lassalle1998}]. In [\onlinecite{bernevig2007}] we found
that the condition $E_0 J^\alpha_\lambda =0$ places severe
restrictions on both the Jack parameter $\alpha$ and on the
partition $\lambda$. We found the following conditions $\alpha < 0$,
$n_0$ $\equiv$ $N-\ell_{\lambda}$ $ > 0$ (non-zero occupancy of the
$m=0$ ``orbital''), as well as
\begin{equation}
N-\ell_{\lambda} + 1 + \alpha(\lambda_{\ell}-1) = 0,
\end{equation}
where $\lambda_{\ell}$ is the smallest (non-zero) part in $\lambda$.
This imposes the following two conditions: (\textit{i}) $\alpha$ is
a negative rational, which we can choose to write as $-(k+1)/(r-1)$,
with $(k+1)$ and $(r-1)$ both positive, and relatively prime;
(\textit{ii}) $\lambda_{\ell}$ = $(r-1)s + 1$, and $n_0$ =
$(k+1)s-1$, where $s > 0$ is a positive integer.  The remaining HW
conditions require that all parts in $\lambda$ have multiplicity
$k$, so that the orbital occupation partition is
$n(\lambda^0_{k,r,s})$ = $[n_0 0^{s(r-1)}k 0^{r-1}k 0^{r-1}k....]$,
(\textit{i.e}, the ($k$,$r$,$N$)-admissibility condition is
satisfied as an equality for orbitals $m$ $\ge$ $\lambda_{\ell}$).

\begin{figure}
\includegraphics[width=3.2in, height=2.1in]{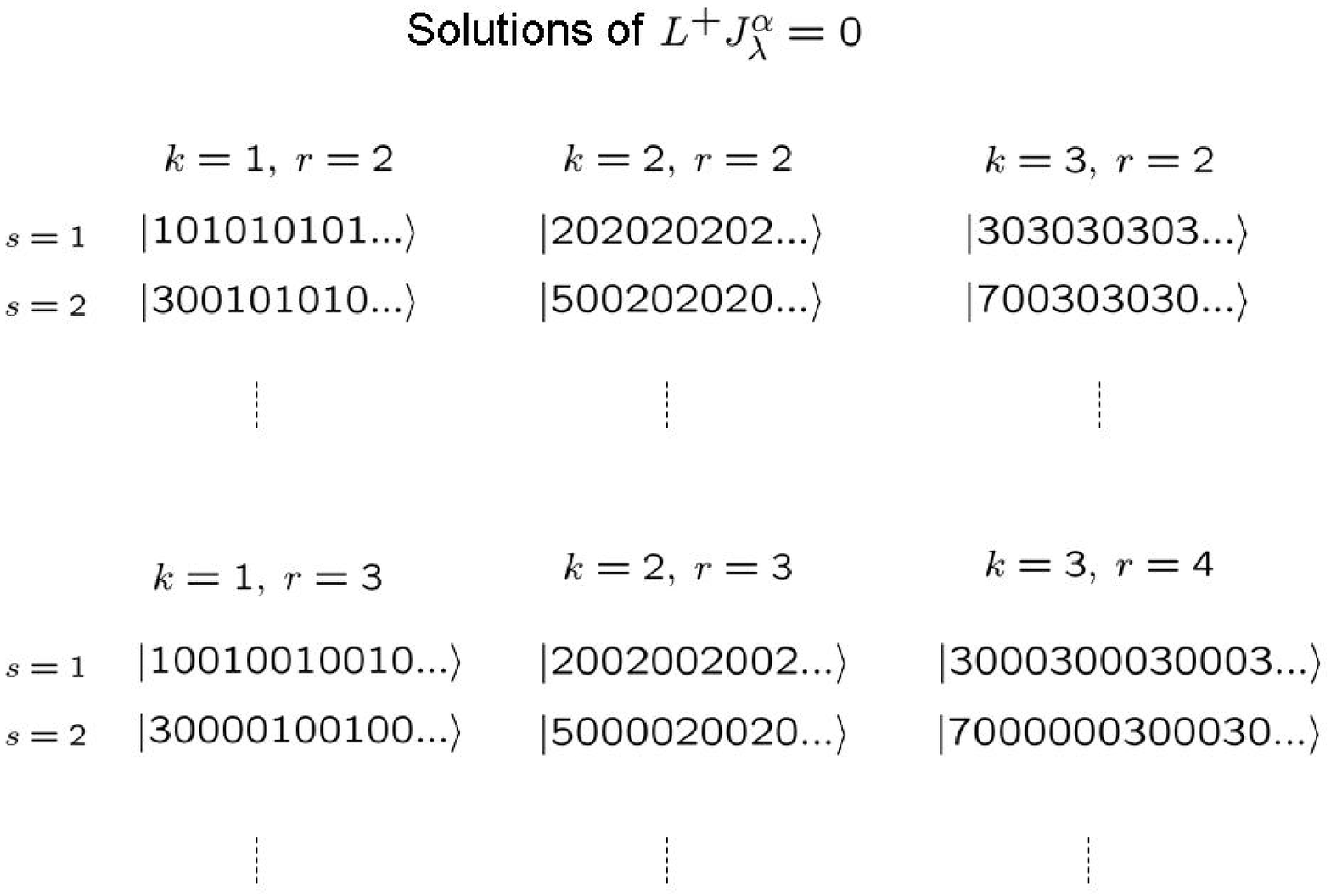}
\caption{Solutions to $L^+ J^\alpha_\lambda=0$ are parametrized by
one integer, $s>0$. Only $s=1$ states are both HW and LW states on
the sphere, and satisfy the clustering property that they vanish as
the $r$'th power of the distance between $k+1$-particles. The $s>1$
states satisfy generalized clustering conditions.}\label{fig2}
\end{figure}

We call these Jacks  HW $(k,r,s,N)$ states
$J^{\alpha_{k,r}}_{\lambda^0_{k,r,s}}$. Non-HW $(k,r,s,N)$ states
with $n_0$ particles in the zeroth orbital can be obtained by
inserting zeroes (holes) in the partition to the right of the
$\lambda_{\ell}$'s orbital. This defines a set of partitions whose
Jacks satisfy the same clustering properties as the HW Jacks
$J^{\alpha_{k,r}}_{\lambda^0_{k,r,s}}$ (see Section III). These
partitions are $\lambda_{k,r,s}:$ $n(\lambda_{k,r,s}) =[n_0
0^{s(r-1)}n(\lambda_{k,r})]$ where $n(\lambda_{k,r})$ is a $(k,r,
N-n_0)$ admissible configuration in the sense of Feigin \emph{et.
al.}. This set of partitions do \emph{not} exhaust the number of
polynomials with the clustering property Eq.(\ref{clustering01}).
The case $s = 1$ gives the generators of the ideals obtained by
Feigin \emph{et. al} and are related to FQH ground states and their
quasihole excitations [\onlinecite{bernevig2007}]. The cases $s> 1$
are new and violate the admissibility conditions of Feigin \emph{el.
al.} (see Fig.\ref{fig2}).

\section{Generalized Clustering Conditions Of Jack Polynomials}

We now present the two generalized clustering conditions satisfied
by the (HW and non-HW) Jacks of the $(k,r,s,N)$ partitions
$J^{\alpha_{k,r}}_{\lambda_{k,r,s}}$.

\subsection{First Clustering Property}

\begin{figure}
\centering
\includegraphics[width=3.3in, height=2.3in]{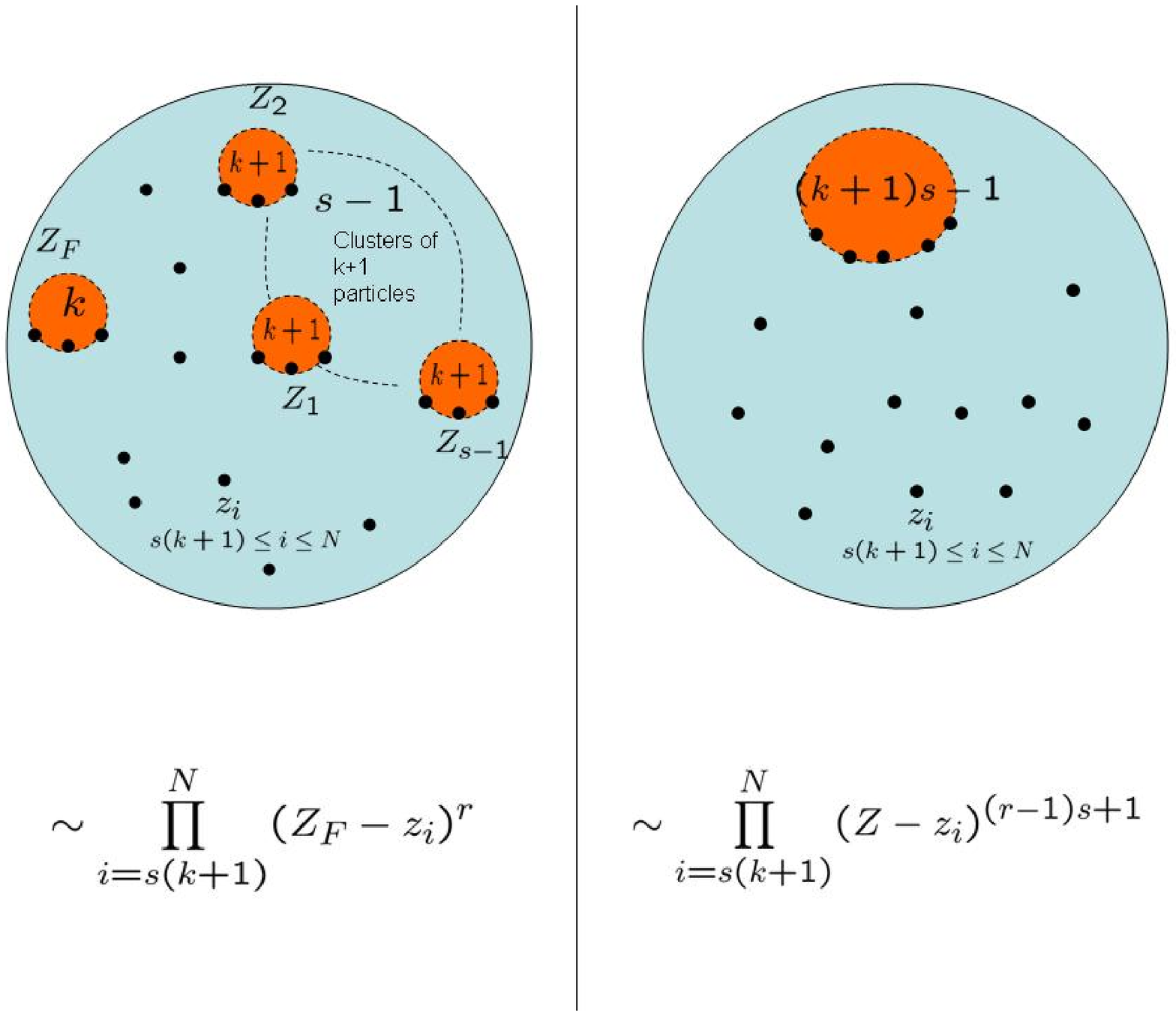}
\caption{Clustering and vanishing conditions of the polynomials
defined by the HW and non-HW Jacks
$J^{\alpha_{k,r}}_{\lambda_{k,r,s}}$}\label{clusteringfig1}
\end{figure}

The Jacks $J^{\alpha_{k,r}}_{\lambda_{k,r,s}}$ allow $s-1$, but not
$s$, different clusters of $k+1$ particles. First form $s-1$
clusters of $k+1$ particles $z_1=...=z_{k+1}(=Z_1)$;
$z_{k+2}=...=z_{2(k+1)}(= Z_2)$; $...$;
 $z_{(s-2)(k+1)+1}=...=z_{(s-1)(k+1)} (= Z_{s-1})$, where the positions of the
 clusters: $Z_1,...,Z_{s-1}$ can be different.
 Then, form a $k$ (\emph{not} $k+1$) particle cluster
 $z_{(s-1)(k+1)+1}=...=z_{s(k+1)-1}(=Z_F)$ (a final  $s$'th cluster of $k+1$
 particles would make the polynomial vanish), see Fig[\ref{clusteringfig1}].
  With the above conditions on
 the particle coordinates, the clustering condition reads:
\begin{equation}
J^{\alpha_{k,r}}_{\lambda_{k,r,s}}(z_1,...,z_N) \propto
\prod_{i=s(k+1)}^N (Z_F-z_i)^r \label{clustering1}
\end{equation}
\noindent Observe that when $s=1$ Eq.(\ref{clustering1}) reduces to
the usual clustering condition satisfied by the $(k,r)$ sequence,
given in [\onlinecite{bernevig2007}].

\subsection{Second Clustering Property} The Jacks $J^{\alpha_{k,r}}_{\lambda_{k,r,s}}$
allow a large cluster of $n_0= (k+1)s-1$ particles at the same
point. As a particular case of Eq.(\ref{clustering1}), they cannot
allow $n_0+1$ particles to come at the same point, as this would
involve the formation of $s$ clusters of $k+1$ particles, which
Eq.(\ref{clustering1}) forbids. Clustering $n_0$ particles at the
same point $z_1=...=z_{(k+1)s-1}=Z$ we find the following property
(see Fig.[\ref{clusteringfig1}]):
\begin{equation}
J^{\alpha_{k,r}}_{\lambda_{k,r,s}}(z_1,...,z_N) \propto
\prod_{i=s(k+1)}^N(Z-z_i)^{(r-1)s+1} \label{clustering10}
\end{equation}
\noindent
 The HW Jacks of partitions $\lambda^0_{k,r,s}$ satisfy an even more
 stringent property; with $z_1=...=z_{(k+1)s-1}=Z$, we find:
\begin{eqnarray}
& J^{\alpha_{k,r}}_{\lambda^0_{k,r,s}}(z_1,...,z_N) =
\prod_{i=s(k+1)}^N(Z-z_i)^{(r-1)s+1}  \times \nonumber \\ & \times
J^{\alpha_{k,r}}_{\lambda^0_{k,r}}(z_{s(k+1)}, z_{s(k+1)+1},...,z_N)
\label{clustering2}
\end{eqnarray}
\noindent where $n(\lambda^0_{k,r})= [k0^{r-1}k0^{r-1}...k]$ is the
maximum density $(k,r,N-n_0)$-admissible partition. For $s=1$,
Eq.(\ref{clustering2}) also reduces to the usual clustering
condition satisfied by the $(k,r)$-admissible sequence
[\onlinecite{bernevig2007}]. We have performed a extensive numerical
checks of the above conjectured clustering conditions. We also note
that the LHS and RHS of Eq.(\ref{clustering2}) match in both total
momentum $M^0$ (total degree of the polynomial):
\begin{eqnarray}
& E_1 J^{\alpha_{k,r}}_{\lambda^0_{k,r,s}}\equiv M^0
J^{\alpha_{k,r}}_{\lambda^0_{k,r,s}} = (N-(k+1)s +1) \times
\nonumber \\ & \times \left[(r-1)s+1 + \frac{1}{2}\frac{r}{k}
(N-(k+1)s - k +1)\right] J^{\alpha_{k,r}}_{\lambda^0_{k,r,s}}
\label{mindegree}
\end{eqnarray}

\begin{figure}
\centering
\includegraphics[width=3.2in, height=2.2in]{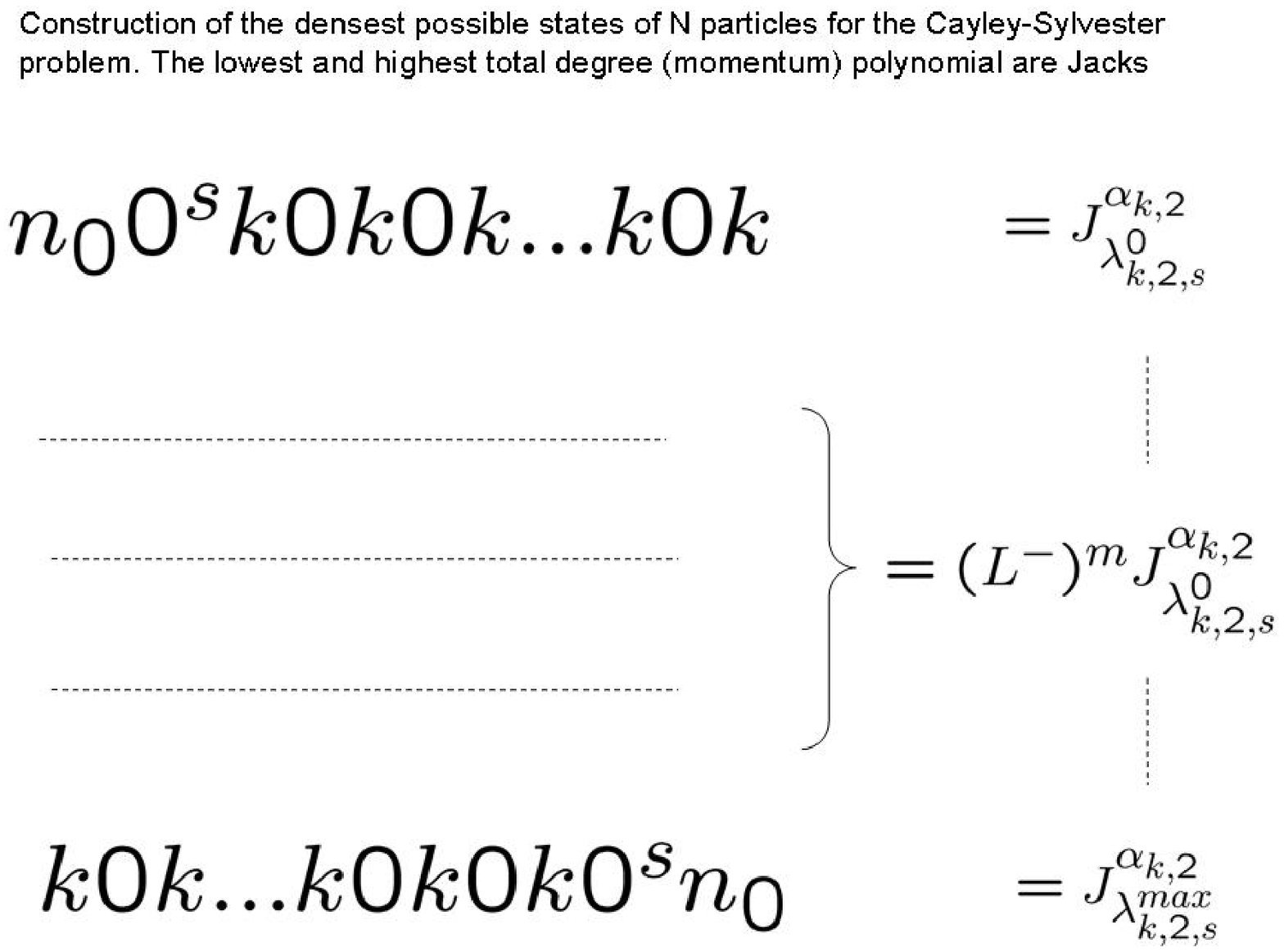}
\caption{The densest polynomials satisfying the clustering
Eq.(\ref{clustering1}) and Eq.(\ref{clustering10}). These are
polynomials of flux $N_\Phi^{0}$ and are obtained by applying $L^-$
on the HW Jack. The LW polynomial is also a Jack, but the
polynomials in between cannot be expanded in terms of only
well-behaved Jack polynomials at $\alpha
=-\frac{k+1}{r-1}$.}\label{densestCayley}
\end{figure}

\noindent and in flux (maximum degree in each variable) $N^0_\Phi$:
\begin{equation}
 N^0_\Phi =\frac{r}{k}\left(N-k - (k+1)(s-1)\right) + (r-1)(s-1).
 \label{minflux}
\end{equation}
\noindent The superscript denotes the fact that we are considering
the momentum and flux of Jack polynomials of HW partitions
$\lambda^0_{k,r,s}$.

\subsection{ Additional Clustering Condition}
We empirically find that the HW Jacks
$J^{\alpha_{k,r}}_{\lambda^0_{k,r,s}}$ satisfy a third type of
clustering which has no correspondence in the $s=1$ case.  Forming
$s-1$ clusters of $2 k+1$ particles together: $z_1=...=z_{2k+1}(=
Z_1)$; $z_{(2k+1) +1} = ...=z_{2(2k+1)}(=Z_2)$;...;$z_{(s-2)(2k+1)
+1} = ...= z_{(s-1)(2k+1)}=(Z_{s-1})$ we find that the HW Jacks
satisfy, up to a numerical proportionality constant, the clustering:
\begin{eqnarray}
& J^{\alpha_{k,r}}_{\lambda^0_{k,r,s}} (z_1,...,z_N) =
\prod_{i<j=1}^{s-1} (Z_i-Z_j)^{k(3 r-2)}  \times \nonumber \\ &
\times  \prod_{i=1}^{s-1}\prod_{l=(s-1)(2k+1)+1}^N (Z_i-z_l)^{2 r
-1} \times \nonumber
\\ & \times J^{\alpha_{k,r}}_{\lambda^0_{k,r}}(z_{(s-1)(2k+1)+1}, ..., z_N)
\label{clustering12}
\end{eqnarray}
\noindent Some slightly tedious algebra proves that the total
momentum and flux match between the LHS and RHS of
Eq.(\ref{clustering12}).

\begin{figure}
\centering
\includegraphics[width=3.4in, height=2.4in]{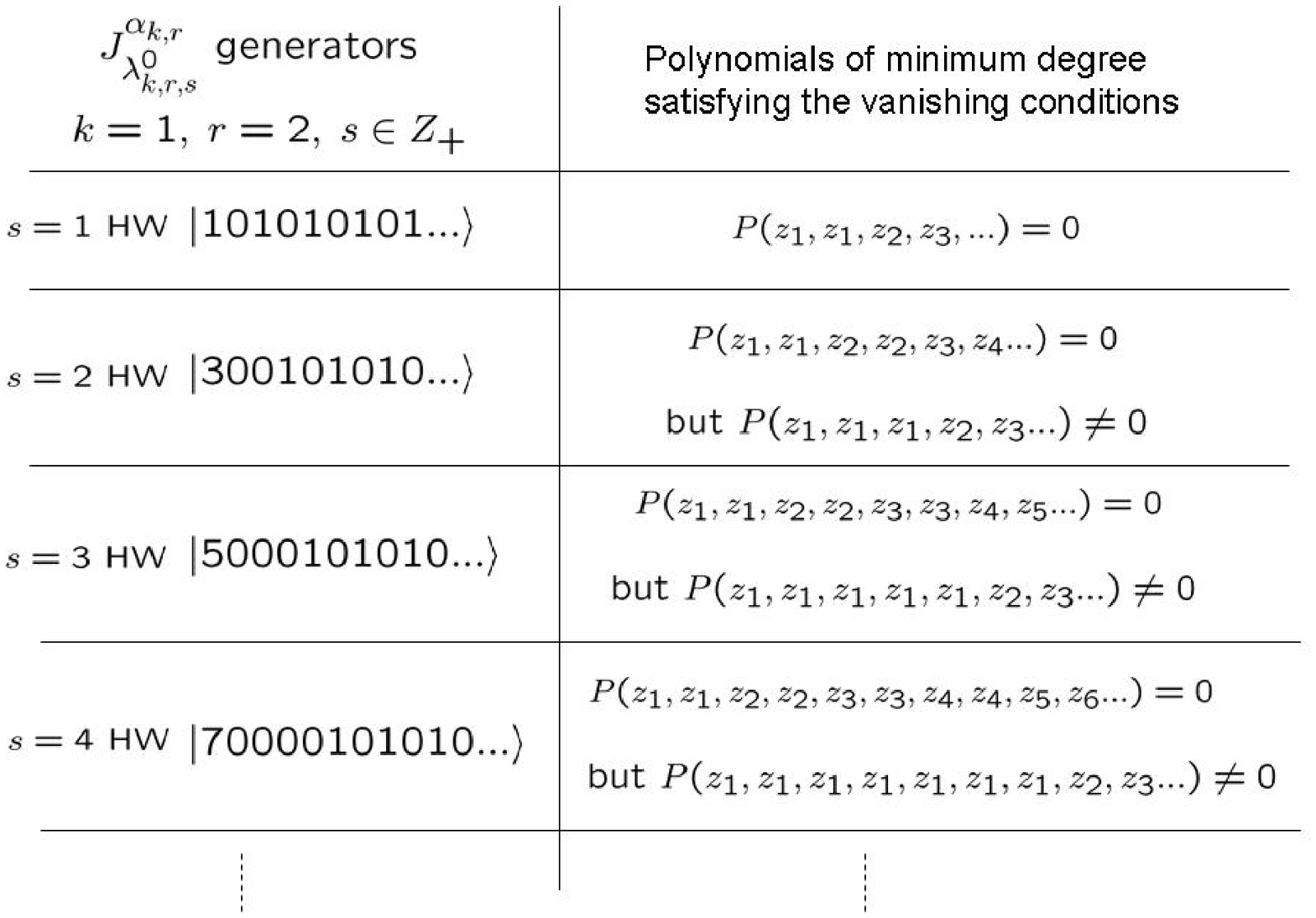}
\caption{Lowest degree polynomials (generators) of the clusterings
$(k,r,s)=(1,2,s)$}\label{clusteringfig2}
\end{figure}

\begin{figure}
\centering
\includegraphics[width=3.4in, height=2.4in]{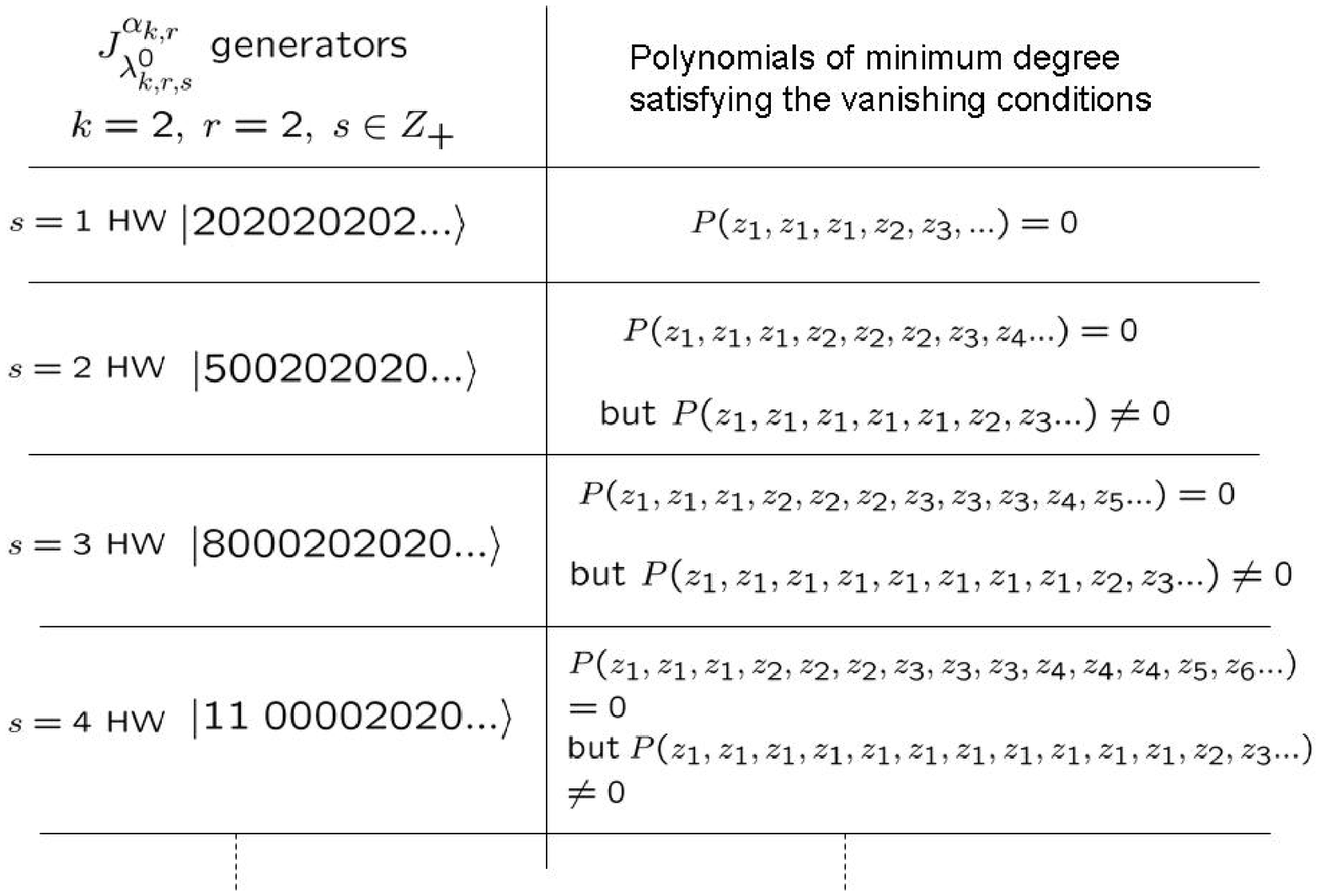}
\caption{Lowest degree polynomials (generators) of the clusterings
$(k,r,s)=(2,2,s)$}\label{clusteringfig3}
\end{figure}

\subsection{Angular Momentum Structure}

The $J^{\alpha_{k,r}}_{\lambda^0_{k,r,s}} (z_1,...,z_N)$ are the HW
states of an angular momentum multiplet of
$l(\lambda^0_{k,r,s})=l^{max}_z$ and $l_z = l^{max}_z,...,-
l^{max}_z$. The LW states of the multiplet $(L^-)^{2 l^{max}_z}
J^{\alpha_{k,r}}_{\lambda^0_{k,r,s}} (z_1,...,z_N)$ are also single
Jack polynomials of the ``symmetric'' partition to
$\lambda^0_{k,r,s}$ in orbital notation (see
Fig.[\ref{densestCayley}]) The value of $l^{max}_z$ is:
\begin{widetext}
\begin{equation}
L^z J^{\alpha_{k,r}}_{\lambda^0_{k,r,s}} \equiv l^{max}_z
J^{\alpha_{k,r}}_{\lambda^0_{k,r,s}}   =
\frac{1}{2}\left[((r-1)s+1)(2(k+1)s -2 -N)+
\frac{r}{k}((k+1)s-1)(N-(k+1)s+
1-k)\right]J^{\alpha_{k,r}}_{\lambda^0_{k,r,s}}
\end{equation}
\end{widetext}
\noindent Most importantly, we find that powers of the operator $L^-
= \sum_i \left( z_i^2 \frac{\partial}{\partial z_i} - N_\Phi
z_i\right)$, acting on $J^{\alpha_{k,r}}_{\lambda^0_{k,r,s}}
(z_1,...,z_N)$ create linearly independent polynomials with the same
vanishing conditions Eq.(\ref{clustering1}), Eq.(\ref{clustering10})
as the HW $J^{\alpha_{k,r}}_{\lambda^0_{k,r,s}} (z_1,...,z_N)$.
There are $2 l^{max}_z +1$ such polynomials.

\section{  $J^{\alpha_{k,r}}_{\lambda^0_{k,r,s}}$: Smallest degree polynomials with generalized clustering }

In the remainder of the paper we focus on the case $r=2$. We
empirically find the following property: given $N$ particles (with
$N>n_0$ for the clustering condition Eq.(\ref{clustering01}) to be
well defined), we find that the $r=2$ HW Jacks
$J^{\alpha_{k,2}}_{\lambda^0_{k,2,s}} (z_1,...,z_N)$ are the
\emph{smallest degree} (smallest momentum $M$ - Eq.(\ref{mindegree})
- and flux $N_\Phi$ - Eq.(\ref{minflux})) polynomials satisfying the
clustering conditions Eq.(\ref{clustering1})
Eq.(\ref{clustering10}). There are \emph{exactly} $2\cdot
l(\lambda^0_{k,2,s})+1$ polynomials in $N$ variables of $N_\Phi^{0}$
(Eq.(\ref{minflux})) and of \emph{unrestricted} total dimensions,
satisfying the clustering Eq.(\ref{clustering1}),
Eq.(\ref{clustering10}). A basis for this ideal is, explicitly:
\begin{equation}
(L^-)^m J^{\alpha_{k,2}}_{\lambda^0_{k,2,s}}; \;\;\;\; m=0,1,...,2
\cdot l(\lambda^0_{k,2,s});
\end{equation}
\noindent We find that $ (L^-)^{2 \cdot
l(\lambda^0_{k,2,s})+1}J^{\alpha_{k,2}}_{\lambda^0_{k,2,s}} =0$. One
can easily understand this counting by looking at the orbital
occupation numbers of the relevant partitions. The occupation number
of the $J^{\alpha_{k,2}}_{\lambda^0_{k,2,s}}$ is $[n_0 0^{s}k
0k0k...k0k]$. This is the lowest weight partition (smallest degree
polynomials) where the clustering conditions Eq.(\ref{clustering1}),
Eq.(\ref{clustering10}) are satisfied. Interpreting this as orbital
occupation number, there exists a ``symmetric" partition
$n(\lambda^{max}_{k,r,s})= [k 0k...k0k0k0^{s}n_0]$. This is the
highest total degree polynomial (bounded by the restriction that
each variable separately has degree at most $N_{\Phi}^{0}$) that
satisfies the vanishing condition Eq.(\ref{clustering01}). It is
also a Jack polynomial $J^{\alpha_{k,r}}_{\lambda^{max}_{k,r,s}}$
(See Fig[\ref{densestCayley}]). Since the $L^-$ operator does not
change the value of $N_\Phi$, maintains the clustering property, and
implements the angular momentum lowering, we then easily count the
polynomials as forming the $l=l_z^{max}$ multiplet.

For the case $k=1, r=2, s=2$, this counting coincides with the
empirical counting observed by Kasatani \emph{et. al.}
[\onlinecite{kasatani2004}].

We remark that the expansion of  $(L^-)^m
J^{\alpha_{k,r}}_{\lambda^0_{k,r,s}}$ in Jack polynomial basis
$J^{\alpha_{k,r}}_\lambda$ contains ill-behaved Jacks which diverge
at the negative $\alpha_{k,r}$ used. However, their coefficients in
the expansion also vanish to give an overall finite contribution.
These are the "modified" Jacks introduced by Kasatani \emph{et. al.}
for the specific $(k,r,s)=(1,2,2)$ case of the problem studied here.
As we reach higher $k$ and $s$ integers, the number of "modified"
Jacks appearing in the expansion of $(L^-)^m
J^{\alpha_{k,r}}_{\lambda^0_{k,r,s}}$ grows large. We therefore
prefer to characterize the basis of these polynomials by the HW Jack
and the polynomials that result from it by successive application of
the $L^-$ operator.

We now relax the constraint $N_\Phi = N_{\Phi}^{0}$ and focus on the
counting of the polynomials satisfying Eq.(\ref{clustering1}) and
Eq.(\ref{clustering10}).

\section{Counting Polynomials}

We want to provide the counting of the number of linearly
independent polynomials in the ideal $F/F_1$. We start by counting
the $s=1$ polynomials. These are related to the admissible
partitions of [\onlinecite{feigin2002}] or the generalized Pauli
principle of [\onlinecite{haldaneUCSB2006},
\onlinecite{bernevig2007}]

\subsection{Counting of $(k,r,s)=(1,r,1)$ Polynomials}

 We first obtain a counting of linearly independent
polynomials in $N$ particle coordinates $z_1,...,z_N$, of total
momentum $M$, with the degree in each coordinate $\le N_\Phi$,
satisfying the condition $P(z_1,z_2,z_3,z_4,..) \sim (z_i-z_j)^r$.
We believe this result was previously known, although we could not
explicitly find it in the literature. From the work of Feigin
\emph{et. al.} [\onlinecite{feigin2002}], this number is equal to
the number of $(k,r)=(1,r)$ admissible partitions of $N_\Phi$
orbitals, related by the squeezing rule [\onlinecite{bernevig2007}]
(so as to keep the partition weight $M$ constant). Call this number
$p_{1,r,1}(N,M,N_\Phi)$. From the theory of partitions
[\onlinecite{andrews1998}], such numbers are most easily obtained
from a generating function $G(q)$, and we can analytically prove
that:
\begin{equation}
p_{1,r,1}(N,M,N_\Phi) = \frac{1}{M!}\frac{\partial^M
G_{1,r,1}(N,N_\Phi,q)}{\partial q^M}\mid_{q=0} \label{p1r1NMNPhi}
\end{equation}
\noindent where the generating function $G(q)$ reads:
\begin{equation}
G_{1,r,1}(N,N_\Phi,q)= \frac{q^{\frac{r}{2}N(N-1)}
\prod_{i=1}^{N_\Phi - r(N-1) + N} (1- q^i)}{\prod_{i=1}^{N} (1-q^i)
\prod_{i=1}^{N_\Phi-r(N-1)} (1-q^i) } \nonumber
\end{equation}

\noindent if $N_\Phi \ge r(N-1)$ and $p_{1,r,1}(N,M,N_\Phi<r(N-1))
=0$. $p_{1,r,1}(N,M,N_\Phi)$ represents a building block for future
results. We have numerically checked, by building the null-space of
polynomials satisfying the clustering condition
$P(z_1,z_1,z_2,z_3,...)=0$, that $p_{1,r,1}(N,M,N_\Phi)$ gives the
right polynomial counting. In the context of FQH, it reproduces the
right counting for quasihole states. For example, it is known that
the Laughlin state with $x$ number of quasiholes has $(N+x)!/(N!
x!)$ independent states and one can numerically check the identity:
\begin{equation}
\left(%
\begin{array}{c}
  N +x\\
  x \\
\end{array}%
\right) =  \sum_{M=\frac{r}{2} N(N-1)}^{\frac{r}{2} N(N-1) + x N}
p_{1,r,1}(N,M,N_\Phi = r(N-1)+x) \nonumber
\end{equation}
\noindent

 We can now provide a formula for the number of polynomials in $N$ variables, of total
dimension $M$, of any maximum power of each coordinate (any
$N_\Phi$), which satisfy the clustering condition $P(z_1,z_2....)
\sim (z_i-z_j)^r$. We
 take $N_\Phi \rightarrow \infty$ to obtain the simpler expression:
\begin{equation}
 p_{1,r,1}(N,M) = \frac{1}{M!}\frac{\partial^M
G_{1,r,1}(N,q)}{\partial q^M}\mid_{q=0} \label{p1r1NM}
\end{equation}
\begin{equation}
G_{1,r,1}(N,q)= \frac{q^{\frac{r}{2}N(N-1)}}{\prod_{i=1}^{N}
(1-q^i)} \nonumber
\end{equation}
\noindent To find out the \emph{total} number of symmetric
polynomials satisfying the vanishing $P(z_1,z_1,z_3, z_4,...,z_N)=0$
we must particularize to the lowest vanishing power possible, $r=2$.
In this case, Eq.(\ref{p1r1NM}) is identical to the formula of
Kasatani \emph{et. al.} [\onlinecite{kasatani2004}], although
Eq.(\ref{p1r1NMNPhi}) represents a more comprehensive counting of
the polynomials as it contains information on the allowed maximum
degree in each variable, $N_\Phi$.

Our aim to conjecture similar expressions for the counting of the
dimension space of the polynomials in the coset space $F/F_1$.

\subsection{Counting of $(k,r,s)=(1,2,s)$ Polynomials}

Using $p_{1,2,1}(N,M,N_\Phi)$, we now obtain the counting of
polynomials in the ideal $F/F_1$ with $k=1,r=2$ and $s>1$. We first
reproduce the result of Kasatani \emph{et. al.}
[\onlinecite{kasatani2004}] which is the case $(k,r,s)=(1,2,2)$ of
our problem.  Kasatani \emph{et. al.} [\onlinecite{kasatani2004}]
obtained the dimension of the linear space of polynomials satisfying
the clustering conditions $P(z_1,z_1, z_2,z_2, z_5, z_6,...,z_N)=0$
and $P(z_1, z_1,z_1,z_4,z_5,...,z_N) \ne 0$, of total dimension $M$,
of any allowed maximum degree in each of the coordinates. We then
derive the general case, which contains information about $N_\Phi$.

 Define $p_{k=1,r=2,s=2}(N,M) =p_{1,2,2}(N,M)  =
p_{1,2,2}(N,M,N_\Phi\le \infty)$ as the number of polynomials in $N$
variables of total momentum (degree) $M$, with any allowed $N_\phi$
($\le M$), satisfying the clustering condition
Eq.(\ref{clustering1}), Eq.(\ref{clustering10}) (with
$k=1,r=2,s=2$). This number can be found as follows: start with the
partition $n(\lambda_{1,2,2})$, of total dimension (weight) $M$
which has $N-1$ particles all pushed maximally to the left of the
orbitals, while the $N$'th particle is pushed as far as needed to
the right so that the polynomial has dimension $M$. This partition
reads $[30010101..101 \underbrace{0...0}_{M-2-N(N-4)} 1]$. Note
that, by $(k,r,s,N)$ admissibility, we cannot push the first $N-1$
particles anymore to the left than they already are. Then
$p_{1,2,2}(N,M)$ is the sum of two terms: First, we can form
$(k,r,s,N) =(1,2,2,N)$ admissible partitions by keeping the
occupancy of the zeroth orbital to be $3$ and by squeezing on the
remainder partition $[00010101..101 \underbrace{0...0}_{M-2-N(N-4)}
1]$ to form all the $(k,r)=(1,2)$-admissible partitions. As
discussed before, this gives Jack polynomials with the same
clustering condition as the HW Jack, and their number is the same as
the number of $(k,r)=(1,2)$ admissible partitions of $N-3$ variables
and total momentum $M-3(N-3)$, i.e.: $p_{1,2,1}(N-3,M-3(N-3))$.
Second, we can form polynomials with the same $(k,r,s,N)=(1,2,2,N)$
clustering by taking some particles out of the zeroth-orbital,
although these now involve divergent Jacks (with compensating
vanishing coefficients). We can form all the polynomials (of
dimension $M$ in $N$ variables, with less than $3$ particles in the
zeroth orbital, and that satisfy the clustering conditions
$(k,r,s,N)=(1,2,2,N)$) by acting with $L^-$ on \emph{all} the
polynomials of dimension $M-1$, that satisfy the same clustering
conditions. This number is then $p_{1,2,2}(N,M-1)$, and we find the
recursion relation:
\begin{equation}
p_{1,2,2}(N,M) = p_{1,2,2}(N,M-1)+p_{1,2,1}(N-3,M-3(N-3))
\label{recurrence1}
\end{equation} \noindent To find the generating function, multiply
Eq.(\ref{recurrence1}) by $q^M$, sum over $M$, re-shift variables in
the sum and obtain:
\begin{equation}
 p_{1,2,2}(N,M) =\frac{1}{M!} \frac{\partial^M
G_{1,2,2}(N,q)}{\partial q^M} \mid_{q=0}
\end{equation}
\begin{equation}
G_{1,2,2}(N,q)=\frac{q^{(N-3)(N-1)}}{(1-q) \prod_{i=1}^{N-3}(1-q^i)}
\nonumber
\end{equation}
\noindent This reproduces a formula obtained by Kasatani \emph{et.
al.} [\onlinecite{kasatani2004}] through different methods.

We now use the same resoning to count the dimension of the ideal
$F/F_1$ with $(k=1,r=2)$ and general $s$. The number of polynomials
of $N$ variables of momentum (total degree) $M$, with unrestricted
$N_\phi$, which satisfy the clustering conditions
Eq.(\ref{clustering1}) and Eq.(\ref{clustering10}) with $k=1, r=2$
and any $s>1$ is $p_{1,2,s}(N,M)$:
\begin{eqnarray}
&p_{1,2,s}(N,M) =\frac{1}{M!} \frac{\partial^M
G_{1,2,s}(N,q)}{\partial q^M} \mid_{q=0}; \nonumber \\ &
G_{1,2,s}(N,q)=\frac{q^{(N-n_0)(N-n_0+s)}}{(1-q)
\prod_{i=1}^{N-n_0}(1-q^i)} \label{p1rsNM}
\end{eqnarray}
\noindent where $n_0 = 2s-1$.

We now introduce information on the maximum degree in each
coordinate separately (flux). Define $p_{k=1,r=2,s}(N,M, N_\Phi)
=p_{1,2,s}(N,M,N_\Phi)$ as the number of polynomials in $N$
variables of total momentum (degree) $M$, with flux $\le N_\phi$,
satisfying the clustering conditions Eq.(\ref{clustering1}) and
Eq.(\ref{clustering10}) with $k=1$ and $s=2$. We briefly present the
reasoning used to conjecture a count of these polynomials. The
smallest dimension and flux correspond to the partition $[n_0
0^s10101...101000000...00]$ (with $n_0=2s-1$). The number of zeroes
on the right is just right to make the total number of orbitals
$N_\Phi+1$.  Some of the orbitals to the right might be unoccupied.
This ``padding'' to the right has the effect of allowing $L^-$ to
move particles up to the right-most orbital. By symmetry in orbital
space, the highest partition corresponds to $[00...000000101...10101
0^s n_0]$ (with total number of orbital $N_\Phi+1$). We can then
immediately see that $p_{1,2,s}(N,M,N_\Phi)=0$ for $M<(s+1)(N-n_0) +
(N-n_0)(N-n_0-1)$ or for $N_\phi <s+1+ 2(N-n_0-1)$. Also,
$p_{1,2,s}(N,M,N_\Phi)=0$ for $M> N N_\Phi - (N-n_0)(N-n_0 +s)$.
There is also an "intermediate" total degree that is important in
the counting, which corresponds to the partition
$[10101...10100000..00n_0]$ of total degree $n_0 N_\Phi +
(N-n_0)(N-n_0 -1)$ when the right-most orbital has been occupied by
the maximum number of particles possible, $n_0$. Then
$p_{1,2,s}(N,M,N_\Phi)$ reads:
\begin{widetext}
\begin{equation}
 p_{1,2,s}(N,M,N_\Phi)= \nonumber
 \end{equation}
\begin{equation}
=  0 \;\;\;\; \text{if} \;\;\;\;  M<(s+1)(N-n_0) + (N-n_0)(N-n_0-1)
\;\;\;\;  or \;\;\;\; N_\phi <s+1+
  2(N-n_0-1) \nonumber
  \end{equation}
\begin{equation}
 = \sum_{i=0}^{M-(s+1)(N-n_0)} p_{1,2,1}(N-n_0,i,N_\Phi - (s+1)) \;\;\;\; \text{if} \;\;\;\;  M\le  n_0 N_\Phi +
(N-n_0)(N-n_0 -1) \nonumber
\end{equation}
\begin{equation}
  =\sum_{i=0}^{N N_{\Phi} - (N-n_0)(s+1)-M} p_{1,2,1}(N-n_0,i,N_\Phi - (s+1)) \;\;\;\; \text{if} \;\;\;\; n_0 N_\Phi +
(N-n_0)(N-n_0 -1) < M \le N N_\Phi - (N-n_0)(N-n_0 +s)\nonumber
\end{equation}
\begin{equation}
  =0 \;\;\;\; \text{if} \;\;\;\; M> N N_\Phi - (N-n_0)(N-n_0 +s) \nonumber
\end{equation}
\end{widetext}
\noindent $p_{1,2,1}(N,M,N_\Phi)$ was explicitly given in a previous
subsection, and $n_0=2 s-1$.

By summing the previous expression over all $M$ we can find the
number of polynomials of $N$ variables, with degree in each variable
$N_\Phi$ and of unrestricted momentum (total degree) $
p_{1,2,s}(N,N_\Phi) = \sum_{M=0}^{\infty}  p_{1,2,s}(N,M,N_\Phi)$,
satisfying the clustering conditions Eq.(\ref{clustering1}) and
Eq.(\ref{clustering10}) with $k=1$ and $s$ arbitrary integer.
However, by applying an empirical rule we observed, based on the
multiplet nature of these polynomials, we find an alternate simpler
formula, which is not obviously equal to $\sum_{M=0}^{\infty}
p_{1,2,s}(N,M,N_\Phi)$; extensive numerical checks have however
confirmed their equivalence:
\begin{widetext}
\begin{equation}
p_{1,2,s}(N,N_\Phi)= \nonumber
\end{equation}
\begin{equation}
=  0 \;\;\;\; \text{if} \;\;\;\;   N_\phi <s+1+
  2(N-n_0-1) \nonumber
\end{equation}
\begin{equation}
=(N N_\Phi - 2(s+1)N + 2 n_0(s+1)+1) \sum_{i=0}^{N N_{\Phi}}
p_{1,2,1}(N-n_0,i,N_\Phi- (s+1)) -2 \sum_{i=0}^{N N_\Phi} i \cdot
p_{1,2,1}(N-n_0,i, N_\Phi-(s+1))
\end{equation}
\end{widetext}
\noindent $p_{1,2,1}(N,M,N_\Phi)$ was explicitly given in subsection
IV A., and $n_0=2 s-1$.

\subsection{Counting of $(k,r,s)=(k,2,1)$ Polynomials}

We now move to the $k>1$ case. We first obtain a count of the
polynomials satisfying the $(k,r)=(k,2)$ statistics, i.e. of the
Read-Rezayi $Z_k$ states. We want to count the number of polynomials
in $N$ variables, of momentum (total degree) $M$ with maximum flux
(maximum degree in each coordinate) $N_\Phi$, that vanish when $k+1$
particles come together. We call this number
$p_{k,2,1}(N,M,N_\Phi)$. As we know [\onlinecite{feigin2002}], this
is equal to the number of $(k,2)$-admissible partitions of weight
$M$, made out of at most $N$ parts, and with $\lambda_1 \le N_\Phi$.
We can derive this by performing a slight modification of a formula
due to Feigin and Loktev [\onlinecite{feigin2000}] (see also Andrews
[\onlinecite{andrews1998}]). $p_{k,2,1}(N,M,N_\Phi)=0$ for $N_\Phi <
\frac{2}{k}(N-k)$ or for $M < \frac{1}{k}N(N-k)$ but otherwise is:
\begin{eqnarray}
&p_{k,2,1}(N,M,N_\Phi)= \frac{1}{M!}\frac{1}{N!}
\frac{\partial^N}{\partial z^N} \frac{\partial^M}{\partial q^M}
\times \nonumber \\ & \times \left(q^{-N} G_{k,2,1}(N_\phi,q,z)
\right)\mid_{q=0,\; z=0} \nonumber
\end{eqnarray}
\noindent where the generating function $G_{k,2,1}(N_\phi,q,z)$ is
[\onlinecite{feigin2000},\onlinecite{andrews1998}]:
\begin{widetext}
\begin{equation}
G_{k,2,1}(N_\phi,q,z) = \sum_{\begin{array}{c}
                                _{m_1,n_1 =0} \\
                                _{m_1+n_1 \le
\frac{N_\Phi+1}{2}} \\
                              \end{array}}^{\frac{N_\Phi+1}{2}} \frac{z^{(k+1)(m_1+n_1)} q^{k(m_1^2-n_1^2 +
n_1(N_\Phi+2)) + \frac{m_1(3m_1-1)}{2}
+n_1(N_\Phi+3-n_1)}}{\prod_{i=1}^{m_1}(1-q^i)(z q^{i+m_1-1}-1)
 \prod_{i=2m_1+1}^{N_\Phi- 2 n_1+1} (1-z q^i)
 \prod_{i=1}^{n_1} (q^i-1)(z q^{N_\Phi -i -n_1+3}-1)   }
\label{genfunckr2s1}
\end{equation}
\end{widetext}
\noindent

We have numerically performed extensive checks of the compatibility
of the formula above, in the case $k=1$, with the simpler expression
of $p_{1,2,1}(N,M,N_\Phi)$ obtained earlier. The formula above also
correctly gives the dimension of the quasihole Hilbert space in the
$Z_k$ parafermions sequence. For one-quasihole, this is known to be:
\begin{equation}
\left(%
\begin{array}{c}
  N/k + k \\
  k \\
\end{array}%
\right) = \sum_{i= \frac{N(N-k)}{k}}^{\frac{N(N-k)}{k} + N}
p_{k,2,1}(N,i,N_\phi=\frac{2}{k}(N-k)+1) \nonumber
\end{equation}
\noindent Extensive numerical checks prove the above identity.
Moreover, for the $k=2$ Read-Moore state with $2$ quasiparticles:
\begin{eqnarray}
&\left(%
\begin{array}{c}
  N/2 + 4 \\
  4 \\
\end{array}%
\right) + \left(%
\begin{array}{c}
  (N-2)/2 + 4 \\
  4 \\
\end{array}%
\right)= \nonumber \\ &=\sum_{i= \frac{N(N-k)}{k}}^{\frac{N(N-k)}{k}
+2 N} p_{k,2,1}(N,i,N_\phi=\frac{2}{k}(N-k)+2) \nonumber
\end{eqnarray}
\noindent

Eq.(\ref{genfunckr2s1}) for $(k,2)$ admissible partitions found by
Feigin and Loktev [\onlinecite{feigin2000}] and prior to them by
Andrews [\onlinecite{andrews1998}] gives the most information
possible about the counting of the Read-Rezayi wavefunctions and
quasiholes. It provides information about the total degree of the
polynomial (multiplet structure), which the usual counting
[\onlinecite{read1999}] of quasiholes does not since it sums over
all the possible total dimensions of the polynomials subject to a
flux $N_\Phi$ upper bound.

\subsection{Counting of the $(k,r,s)=(k,2,s)$ Polynomials}

Using $p_{k,2,1}(N,M,N_\Phi)$ we obtain the counting of polynomials
in the $F/F_1$ ideal with arbitrary $k$ and $s$. Following a line of
reasoning similar to the one used in the $k=1$ case, we find the
number $p_{k,2,s}(N,M,N_\Phi)$ of polynomials in $N$ variables, of
momentum (total degree) $M$, of flux $N_\Phi$ that have the
clustering conditions Eq.(\ref{clustering1}) and
Eq.(\ref{clustering10}) for general $k$ and $s>1$ integers reads:
\begin{widetext}
\begin{equation}
p_{k,2,s}(N,M,N_\Phi) =\nonumber
\end{equation}
\begin{equation}
=0; \;\;\;\; \text{if} \;\;\;\; M <(s+1)\cdot (N-n_0) + \frac{1}{k}
(N-n_0)(N-n_0-k) \;\;\;\; or \;\;\;\; N_\Phi < s+1 +
\frac{2}{k}(N-n_0 -k) \nonumber
\end{equation}
\begin{equation}
=\sum_{i=0}^{M-(s+1)(N-n_0)} p_{k,2,1}(N-n_0,i, N_\Phi-(s+1)) ;
\;\;\;\; \text{if} \;\;\;\; 0\le M \le n_0 N_\Phi +
\frac{1}{k}(N-n_0)(N-n_0-k) \nonumber
\end{equation}
\begin{equation}
=\sum_{i=0}^{N N_\Phi -(N-n_0)(s+1) -M} p_{k,2,1}(N-n_0, i, N_\Phi
-(s+1)); \;\;\;\; n_0 N_\Phi + \frac{1}{k}(N-n_0)(N-n_0 - k) < M \le
N N_\Phi - (N-n_0)(s+ \frac{N-n_0}{k})
\end{equation}
\begin{equation}
=0 \;\;\;\; \text{if}  \;\;\;\; M > N N_\Phi - (N-n_0)(s+
\frac{N-n_0}{k})
\end{equation}
\end{widetext}
\noindent where $n_0=(k+1)s -1$

By summing the previous expression over all $M$ we can find the
number of polynomials of $N$ variables, with flux $N_\Phi$ and of
unrestricted momentum (total degree) $ p_{k,2,s}(N,N_\Phi) =
\sum_{M=0}^{\infty}  p_{k,2,s}(N,M,N_\Phi)$, satisfying the
clustering conditions Eq.(\ref{clustering1}) and
Eq.(\ref{clustering10}) with $k$ and $s$ arbitrary integers.
However, by using the angular momentum multiplet structure of these
polynomials, we find an alternate formula which is not obviously
equal to $\sum_{M=0}^{\infty} p_{k,2,s}(N,M,N_\Phi)$; extensive
numerical checks have however confirmed their equivalence:
\begin{widetext}
\begin{equation}
p_{k,2,s}(N,N_\Phi)= \nonumber
\end{equation}
\begin{equation}
=  0 \;\;\;\; \text{if} \;\;\;\;   N_\phi <s+1+
  \frac{2}{k} (N-n_0-k) \nonumber
\end{equation}
\begin{equation}
=(N N_\Phi - 2(s+1)N + 2 n_0(s+1)+1) \sum_{i=0}^{N N_{\Phi}}
p_{k,2,1}(N-n_0,i,N_\Phi- (s+1)) -2 \sum_{i=0}^{N N_\Phi} i \cdot
p_{k,2,1}(N-n_0,i, N_\Phi-(s+1))
\end{equation}
\end{widetext}
\noindent $p_{k,2,1}(N,M,N_\Phi)$ is given in VC., and $n_0=(k+1)
s-1$.

\section{Subideals of the Cayley-Sylvester Problem}

So far we have focused on the
 ideal $F/F_1$. We can systematically characterize the ideal $F$ of
 polynomials $P(z_1,z_2,...,z_N)$ which vanish when we form $s$ clusters of $k+1$
particles in the following way: let the \emph{sub-ideals} $F_i$ be
the polynomials
 that satisfy (Eq.(\ref{clustering01})) but that also vanish when $(k+1)s-i$ particles are brought
at the same point. Then $F=\bigcup_{i=0}^{k+1} F_i/F_{i+1}$, where
$F_0=F$ and $F_{k+1}$ is the ideal of polynomials that vanish when
$s-1$ clusters of $k+1$ particles are formed ($F_{k+2} =
\emptyset$). Hence the polynomial ideal $F_i$ is defined by the two
clustering conditions:
\begin{widetext}
\begin{equation}
P(z_1=...=z_{k+1},z_{k+2}=...=z_{2(k+1)},...,z_{(s-1)(k+1)+1}=...=z_{s(k+1)},
z_{s(k+1)+1},z_{s(k+1)+2},...,z_N) =0
\end{equation}
\noindent and
\begin{equation}
P(z_1=...=z_{s(k+1)-i}, z_{s(k+1)-i+1},z_{s(k+1)-i+2},...,z_N) = 0
\end{equation}
\end{widetext}
\noindent  We have not found the generators for the $F_i/F_{i+1}$
ideals, nor have were we able to find their counting rules for the
general case. However, we have solved the problem for several
specific cases which we present below.

\subsection{Subideals of the $(k,r,s) = (k,2,2)$}

\begin{figure}
\centering
\includegraphics[width=3.4in, height=2.4in]{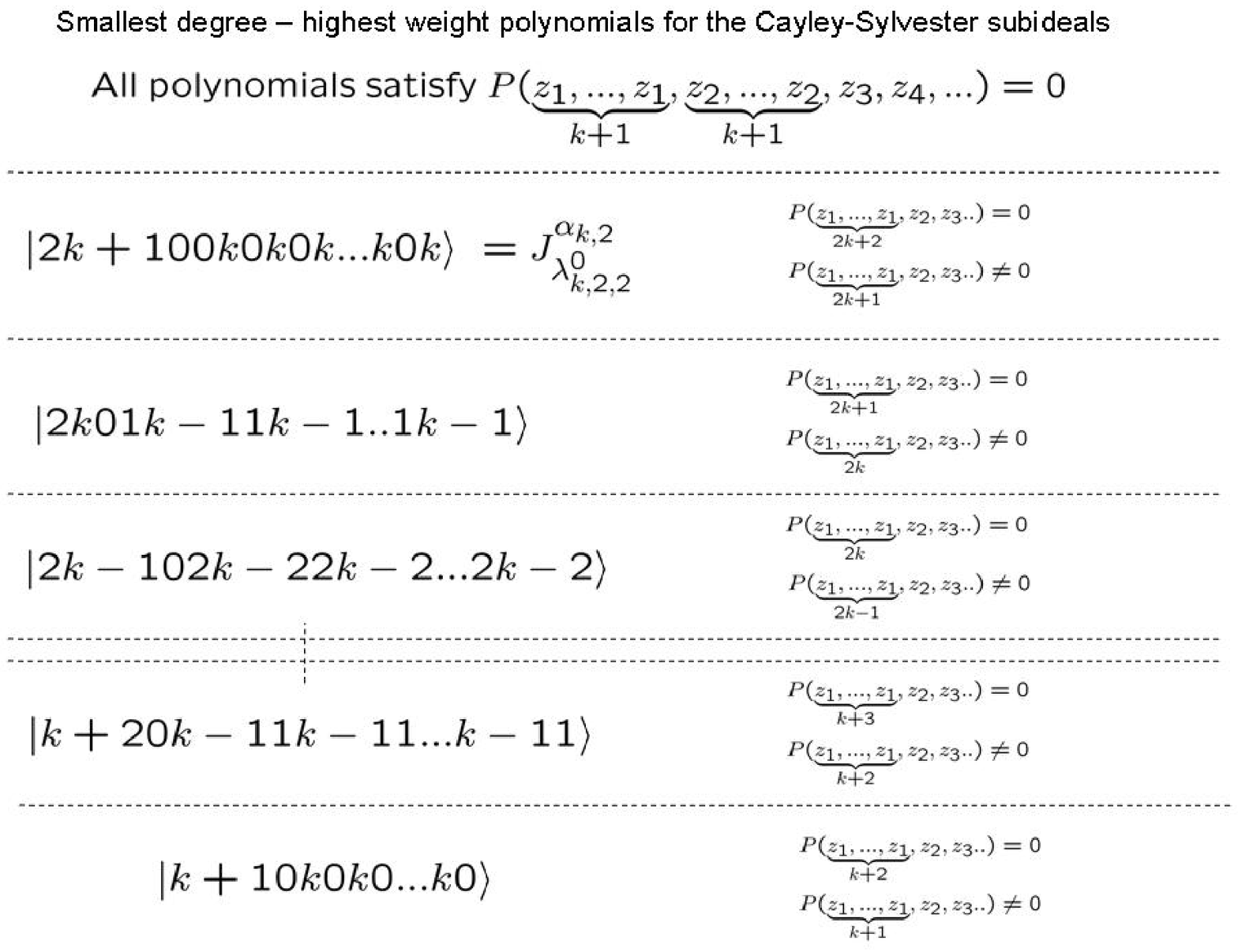}
\caption{Cayley-Sylvester subideals for polynomials satisfying $P(
z_1,...,z_1, z_2,...,z_2, z_3, z_4,...)=0$ }
\label{CayleySylvesterSubideals2}
\end{figure}

We now give the partitions for the generators (smallest degree
highest weight polynomials) for the subideals $F_i/F_{i+1}$ for the
infinite series $(k,r,s)=(k,2,2)$. The first smallest degree
polynomials that vanish when $2$ distinct clusters of $k+1$
particles are formed, but does not vanish when one large cluster of
$2k+1$ particles is formed, is dominated by the root partition:
\begin{eqnarray}
& |2k+1 00 k0k0k0k... k0k \rangle : \nonumber \\ &
P(\underbrace{z_1,...,z_1}_{k+1}, \underbrace{z_2,...,z_2}_{k+1},
z_3, z_4,....) =0 \nonumber \\ & P(\underbrace{z_1,...,z_1}_{2k+1},
z_2, z_3,....) \ne 0
\end{eqnarray}
\noindent The polynomial above, as well as the ones we introduce
below, can be written as a linear combinations of monomials of
partitions dominated by the root partition above, with coefficients
that are \emph{uniquely} defined by the HW and clustering
conditions. These are of course, the Jacks. Then the smallest degree
polynomial that vanishes when either $2$ distinct clusters of $k+1$
particles are formed or when a large single cluster of $2k+1$
particles is formed, but does not vanish when one large cluster of
$2k$ particles if formed, is dominated by the partition ( see
Fig[\ref{CayleySylvesterSubideals2}]):
\begin{eqnarray}
&|2k 01 k-11k-11k-1... 1k-1 \rangle : \nonumber \\ &
P(\underbrace{z_1,...,z_1}_{k+1}, \underbrace{z_2,...,z_2}_{k+1},
z_3, z_4,...) =0  \\ & P(\underbrace{z_1,...,z_1}_{2k+1}, z_2,
z_3,...) = 0, \;\; P(\underbrace{z_1,...,z_1}_{2k}, z_2, z_3...) \ne
0 ; \nonumber
\end{eqnarray}
\noindent The smallest degree polynomial that vanishes when either
$2$ distinct clusters of $k+1$ particles are formed or when a large
single cluster of $2k$ particles is formed but does not vanish when
one large cluster of $2k-1$ particles if formed, is dominated by the
partition ( see Fig[\ref{CayleySylvesterSubideals2}]):
\begin{eqnarray}
&|2k-1 02 k-22k-22k-2... 2k-2 \rangle : \nonumber \\ &
P(\underbrace{z_1,...,z_1}_{k+1}, \underbrace{z_2,...,z_2}_{k+1},
z_3, z_4,...) =0  \\ & P(\underbrace{z_1,...,z_1}_{2k}, z_2,
z_3,...) = 0, \;\; P(\underbrace{z_1,...,z_1}_{2k-1}, z_2, z_3,...)
\ne 0; \nonumber
\end{eqnarray}
\noindent and so on until. At last, the smallest degree polynomial
that vanishes when either $2$ distinct clusters of $k+1$ particles
are formed or when a single cluster of $k+2$ particles is formed but
does not vanish when one cluster of $k+1$ particles if formed, is
dominated by the partition ( see
Fig[\ref{CayleySylvesterSubideals2}]):
\begin{eqnarray}
& |k+1 0k 0k0k... k0 \rangle: \nonumber \\ &
P(\underbrace{z_1,...,z_1}_{k+1}, \underbrace{z_2,...,z_2}_{k+1},
z_3, z_4,....) =0 \\ & P(\underbrace{z_1,...,z_1}_{k+2}, z_2,
z_3,...) = 0, \;\; P(\underbrace{z_1,...,z_1}_{k+1}, z_2, z_3,...)
\ne 0; \nonumber \label{lastsubideal}
\end{eqnarray}

The polynomials of the last sub-ideal, Eq.(\ref{lastsubideal}) are
related to the quasiparticle excitations of abelian and non-abelian
FQH states [\onlinecite{uslater}]. They perform well under
$k+2$-body repulsive interactions. We can also find the smallest
weight partitions of polynomials that vanish when $s$ clusters of
$k+1$ particles come together and when $k+2$ particles come
together, but do not vanish when $k+1$ particles form a cluster:
\begin{eqnarray}
&|\underbrace{k+10k+10...k+10}_{s-1}k0k0k...0k \rangle: \nonumber \\
& P(\underbrace{z_1,...,z_1}_{k+1},
...,\underbrace{z_s,...,z_s}_{k+1}, z_{s+1},z_{s+2},...)=0
\\ & P(\underbrace{z_1,...,z_1}_{k+2}, z_2, z_3,...) =0,\;\;
P(\underbrace{z_1,...,z_1}_{k+1}, z_2, z_3,...) \ne 0 \nonumber
\end{eqnarray}

\subsection{Counting of the $F_k/F_{k+1}$ Subideal}

The counting of dimension of the subideals above is a rather
difficult (but tractable) problem. The ``easy'' exceptions are the
first subideal, whose generator is $|2k+1 00k0k0k...k0k\rangle$ and
whose counting formulas we have already conjectured in the body of
this manuscript, and the last subideal whose generator is
$|k+10k0k0k...k0k\rangle$, and whose counting formula we conjecture
below. The number $p^{F_k/F_{k+1}}_{k,2,2}(N,M,N_\Phi)$ of
polynomials,
 satisfying the clusterings of Eq.(\ref{lastsubideal}), of $N$ variables, of
momentum (total degree $M$) and of flux (maximum separate degree in
any variable) $N_\Phi$  is:
\begin{widetext}
\begin{equation}
p^{F_k/F_{k+1}}_{k,2,2}(N,M,N_\Phi) =\nonumber
\end{equation}
\begin{equation}
=0; \;\;\;\; \text{if} \;\;\;\; M <2\cdot (N-n_0) + \frac{1}{k}
(N-n_0)(N-n_0-k) \;\;\;\; or \;\;\;\; N_\Phi < 2 + \frac{2}{k}(N-n_0
-k) \nonumber
\end{equation}
\begin{equation}
=\sum_{i=0}^{M-2(N-n_0)} p_{k,2,1}(N-n_0,i, N_\Phi-2) ; \;\;\;\;
\text{if} \;\;\;\; 0\le M \le n_0 N_\Phi +
\frac{1}{k}(N-n_0)(N-n_0-k) \nonumber
\end{equation}
\begin{equation}
=\sum_{i=0}^{N N_\Phi -(N-n_0)2 -M} p_{k,2,1}(N-n_0, i, N_\Phi -2);
\;\;\;\; n_0 N_\Phi + \frac{1}{k}(N-n_0)(N-n_0 - k) < M \le N N_\Phi
- (N-n_0)(1+ \frac{N-n_0}{k})
\end{equation}
\begin{equation}
=0 \;\;\;\; \text{if}  \;\;\;\; M > N N_\Phi - (N-n_0)(1+
\frac{N-n_0}{k})
\end{equation}
\end{widetext}
\noindent where $n_0=k+1$.

By summing the previous expression over all $M$ we can find the
number of polynomials of $N$ variables, with degree in each variable
at most $N_\Phi$ and of unrestricted momentum (total degree) $
p^{F_k/F_{k+1}}_{k,2,2}(N,N_\Phi) = \sum_{M=0}^{\infty}
p^{F_k/F_{k+1}}_{k,2,2}(N,M,N_\Phi)$. By applying a rule based on
the multiplet nature of these polynomials, we find an alternate
formula which is not obviously equal to $\sum_{M=0}^{\infty}
p^{F_k/F_{k+1}}_{k,2,2}(N,M,N_\Phi)$; extensive numerical checks
have however confirmed their equivalence:
\begin{widetext}
\begin{equation}
p^{subideal}_{k,2,2}(N,N_\Phi)= \nonumber
\end{equation}
\begin{equation}
=  0 \;\;\;\; \text{if} \;\;\;\;   N_\phi <2+
  \frac{2}{k} (N-n_0-k) \nonumber
\end{equation}
\begin{equation}
=(N N_\Phi - 4 N + 4 n_0+1) \sum_{i=0}^{N N_{\Phi}}
p_{k,2,1}(N-n_0,i,N_\Phi- 2) -2 \sum_{i=0}^{N N_\Phi} i \cdot
p_{k,2,1}(N-n_0,i, N_\Phi-2)
\end{equation}
\end{widetext}
\noindent where $n_0=k+1$.

\section{Conclusions}

In this paper we have made several new conjectures about the
behavior of Jack polynomials at negative Jack parameter $\alpha$. By
applying a HW condition, we find that the $(k,r)$-admissible
partitions of Feigin \emph{et.al} [\onlinecite{feigin2002}] do not
exhaust the space of partitions for which the Jack polynomials are
well-behaved. We find a new infinite series of Jacks, described by a
positive integer $s$ that vanish when $s$ distinct clusters of $k+1$
particles are formed, but do not vanish when a large cluster of
$s(k+1)-1$ particles is formed. We conjecture an empirical counting
of polynomials with such clustering properties. We also find the
dominant partitions and counting of polynomials that vanish when
either $s$ distinct clusters of $k+1$ particles are formed or a
cluster of $k+2$ particles are formed, but do not vanish when a
large cluster of $k+1$ particles is formed. These results will be of
physical use in the description of the quasiparticle excitations of
the abelian and non-abelian Fractional Quantum Hall states
[\onlinecite{uslater}].

\end{document}